\documentclass[runningheads]{llncs}
\usepackage{graphicx}

\usepackage{amssymb}

\usepackage{hyperref}
\hypersetup{
	colorlinks=true,       
	linkcolor=blue,          
	citecolor=blue,        
	filecolor=blue,      
	urlcolor=blue           
}

\newcommand{\GR}{G_{\mathrm{R}}}
\newcommand{\Grr}{G_{\mathrm{rr}}}
\newcommand{\Gpr}{G_{\mathrm{pr}}}
\newcommand{\Gqr}{G_{\mathrm{qr}}}
\newcommand{\Gqrd}{G_{\mathrm{qrd}}}
\newcommand{\Gqrnd}{G_{\mathrm{qrnd}}}
\newcommand{\WR}{W_{\mathrm{R}}}
\newcommand{\Wrr}{W_{\mathrm{rr}}}
\newcommand{\Wpr}{W_{\mathrm{pr}}}
\newcommand{\Wqr}{W_{\mathrm{qr}}}
\newcommand{\Wqrnd}{W_{\mathrm{qrnd}}}

\begin{document}
\title{Contagion in Bitcoin networks}
%
%
\author{C\'elestin Coquid\'e\inst{1} \and
Jos\'e Lages\inst{1}
\and
Dima L. Shepelyansky\inst{2}
}
\authorrunning{C. Coquid\'e et al.}
%
\institute{Institut UTINAM, OSU THETA, 
Universit\'e de Bourgogne Franche-Comt\'e, CNRS, Besan\c con, France\\
\email{\{celestin.coquide,jose.lages\}@utinam.cnrs.fr}\\
\and
Laboratoire de Physique Th\'eorique, IRSAMC, 
Universit\'e de Toulouse, CNRS, UPS, 31062 Toulouse, France\\
\email{dima@irsamc.ups-tlse.fr}\\
}
\maketitle              
\begin{abstract}
We construct the Google matrices of bitcoin transactions 
for all year quarters 
during the period of January 11,  2009 till 
April 10, 2013. During the last quarters the network size contains
about 6 million users (nodes) with about 150 million transactions.
From PageRank and CheiRank probabilities, analogous to 
trade import and export,
we determine the dimensionless trade balance of each user and
model the contagion propagation on the network
assuming that a user goes bankrupt if its balance exceeds
a certain dimensionless threshold $\kappa$. 
We find that the phase transition takes place 
for $\kappa < \kappa_c \approx 0.1$ with almost 
all users going bankrupt. For $\kappa > 0.55$
almost all users remain safe. We find that even 
on a distance from 
the critical threshold $\kappa_c$ the top PageRank and CheiRank 
users, as a house of cards, rapidly drop to the bankruptcy.
We attribute this effect to strong interconnections between
these top users which we determine
with the reduced Google matrix algorithm.
This algorithm allows to establish efficiently 
the direct and indirect interactions
between top PageRank users. We argue that this study 
models the contagion on real financial networks.

\end{abstract}

\setlength{\tabcolsep}{3pt}

\section{Introduction}
%
%
%
%
%
%
The financial crisis of 2007-2008 produced an enormous impact
on financial, social and political 
levels for many world countries (see e.g. \cite{fcrisiswiki,fcrisisguardian}).
After this crisis the importance of contagion in financial
networks gained a practical importance and
generated serious academic research
with various models proposed for the description
of this phenomenon (see e.g. Reviews \cite{gai,jackson}).
The interbank contagion is of especial interest
due to possible vulnerability of banks during periods of crisis
(see e.g. \cite{goetzcontag,meller}). The bank networks 
have relatively small size with about $N \approx 6000$ 
bank units (nodes)
for the whole US Federal Reserve \cite{soramaki}
and about $N \approx 2000$ for bank units of Germany \cite{craig2014}.
However, the access to these bank networks
is highly protected that makes essentially
forbidden any academic research of real bank networks.

However, at present the transactions in cryptocurrency
are open to public and the analysis of the related 
networks are accessible for academic research.
The first cryptocurrency is bitcoin 
launched in 2008 \cite{nakamoto}. The first steps in the network
analysis of bitcoin transactions are reported in \cite{shamir,biryukov}
and overview of bitcoin system development is given in \cite{bitcoinscience}.
The Google matrix analysis of the bitcoin network (BCN) 
has been pushed forward in \cite{efsbitcoin}  demonstrating that
 the main part of wealth of the network is captured by a small fraction of users.
The Google matrix $G$ describes the Markov transitions on directed
networks and is at the foundations of Google search engine \cite{brin,meyer}.
It finds also useful applications for variety of directed networks 
describe in \cite{rmp2015}. The ranking of network nodes
is based on the PageRank and CheiRank probabilities of  $G$ matrix
which are on average proportional to the number of ingoing and outgoing
links being similar to import and export in the world trade network 
\cite{wtn1,wtn2}. We use these probabilities to determine the balance
of each user (node) of bitcoin network and model the 
contagion of users using the real data of bitcoin transactions from
January 11, 2009 till April 10, 2013. We also analyze the direct and hidden (indirect)
links between top PageRank users of BCN using the recently developed reduced Google matrix
(REGOMAX) algorithm \cite{greduced,politwiki,wtn3,wto}.

\begin{table}[ht!]
	\caption{List of Bitcoin transfer networks. The BC$yy$Q$q$ Bitcoin network corresponds to transactions between active users during the $q$th quarter of year 20$yy$. $N$ is the number of users and $N_{l}$ is the total amount of transactions in the corresponding quarter.}
	\centering
	\begin{tabular}{|lrr|lrr|lrr|}
		\hline
		Network&$N$&$N_{l}$&Network&$N$&$N_{l}$&Network&$N$&$N_{l}$\\
		\hline
		BC10Q3&37818&57437&BC11Q3&1546877&2857232&BC12Q3&3742174&8381654\\
		BC10Q4&70987&111015&BC11Q4&1884918&3635927&BC12Q4&4671604&11258315\\
		BC11Q1&204398&333268&BC12Q1&2186107&4395611&BC13Q1&5997717&15205087\\
		BC11Q2&696948&1328505&BC12Q2&2645039&5655802&BC13Q2&6297009&16056427\\
		\hline
	\end{tabular}
	\label{tab1}
\end{table}

\section{Datasets, algorithms and methods}
We use the bitcoin transaction data
described in \cite{efsbitcoin}. However, there the network was constructed from 
the transactions performed from the very beginning till a given moment of time
(bounded by April 2013). Instead, here we construct the network
only for time slices formed by quarters of calendar year.
Thus we obtain 12 networks with $N$ users and $N_l$ directed links
for each quarter given in Table~\ref{tab1}. We present our main results for BC13Q1.

The Google matrix $G$ of BCN is constructed in the standard way as it is described in detail
in \cite{efsbitcoin}. Thus all bitcoin transactions from a given user (node) to
other users are normalized to unity, the columns of dangling nodes with zero
transactions are replaced by a column with all elements being $1/N$.
This forms $S$ matrix of Markov transitions which is multiplied by the damping factor 
$\alpha=0.85$ so that finally $G=\alpha S +(1-\alpha)E/N$ where the matrix $E$
has all elements being unity.  We also construct the matrix $G^*$ for the inverted
direction of transactions and then following the above procedure for $G$.
The PageRank vector $P$ is the right eigenvector of $G$,
$G P = \lambda P$, with the largest eigenvalue $\lambda=1$ ($\sum_j P(j)=1$).
Each component $P_u$ with $u\in\{u_1,u_2,\dots,u_N\}$ is positive and gives the probability to find a random surfer at the given node $u$ (user $u$). In a similar way the CheiRank vector $P^*$ is defined as the right eigenvector of $G^*$ with eigenvalue $\lambda^*=1$, i.e., $G^* P^* = P^*$. Each component $P^*_u$ of $P^*$  gives the CheiRank probability to find a random surfer on the given node $u$ (user $u$) of the network with inverted direction of links (see \cite{rmp2015,cheirank}).
We order all users $\{u_1,u_2,\dots,u_N\}$ by decreasing PageRank probability $P_u$. We define the PageRank index $K$ such as we assign $K=1$ to user $u$ with the maximal $P_u$, then we assign $K=2$ to the user with the second most important PageRank probability, and so on ..., we assign $K=N$ to the user with the lowest PageRank probability. Similarly we define the CheiRank indexes $K^*=1,2,\dots,N$ using CheiRank probabilities $\{P_{u_1}^*,P_{u_2}^*,\dots,P_{u_N}^*\}$. $K^*=1$ ($K^*=N$) is assigned to user with the maximal (minimal) CheiRank probability.

The reduced Google matrix $\GR$ is constructed for a selected subset of
$N_r$ nodes. 
The construction is based on methods of scattering theory 
used in different fields including mesoscopic and nuclear physics, and  
quantum chaos. It describes, in a matrix of size $N_r \times N_r$,
the full contribution of direct and indirect pathways,  happening 
in the global network of $N$ nodes,  between  $N_r$ selected nodes of interest. 
The PageRank probabilities of the $N_r$ nodes are the same 
as for the global network with $N$ nodes,
up to a constant factor taking into account that 
the sum of PageRank probabilities over $N_r$
nodes is unity. The $(i,j)$-element of $\GR$ 
can be viewed as the probability for a random seller (surfer) starting at 
node $j$ to arrive in node $i$ using direct and indirect interactions. 
Indirect interactions describes  pathways composed in part of nodes different 
from the $N_r$ ones of interest.    
The computation steps of $\GR$ offer 
a decomposition into matrices that clearly distinguish 
direct from indirect interactions,
$\GR = \Grr + \Gpr + \Gqr$ \cite{politwiki}.
Here $\Grr$ is generated  by the direct links between selected 
$N_r$ nodes in the global $G$ matrix with $N$ nodes. The matrix 
 $\Gpr$ is usually rather close to 
the matrix in which each column is given by 
the PageRank vector $P_r$. 
Due to that $\Gpr$ does not bring much information about direct 
and indirect links between selected nodes.
The interesting role is played by $\Gqr$. It takes 
into account all indirect links between
selected nodes appearing due to multiple pathways via 
the $N$ global network nodes (see~\cite{greduced,politwiki}).
The matrix  $\Gqr = \Gqrd + \Gqrnd$ has diagonal ($\Gqrd$)
and non-diagonal ($\Gqrnd$) parts where $\Gqrnd$
 describes indirect interactions between nodes.
The explicit mathematical formulas  and numerical computation 
methods of all three matrix components of $\GR$ are given 
in \cite{greduced,politwiki,wtn3,wto}. 

Following \cite{wtn2,wtn3,wto}, we remind that the PageRank (CheiRank) probability of a user $u$ is related to its ability to buy (sell) bitcoins, we therefore determine the balance of a given user as
$B_u = (P^*(u) - P(u))/(P^*(u) + P(u))$.
We consider that a user $u$ goes to bankruptcy if $B_u \leq - \kappa$. If it is the case the user $u$  ingoing flow of bitcoins is stopped. This is analogous
to the world trade case when countries with unbalanced trade stop their import
in case of crisis \cite{wtn1,wtn2}. Here $\kappa$ has the meaning of 
bankruptcy or crisis threshold. Thus the contagion model is defined as follows:
at iteration $\tau$, the PageRank and CheiRank probabilities are computed taking into account that all ingoing bitcoin transactions to users went to bankruptcy at previous iterations are stopped (i.e., these transactions are set to zero). Using these new PageRank and CheiRank probabilities we compute again the balance of each user, determining which additional users went to bankruptcy at iteration $\tau$. Initially at the first iteration, $\tau=1$, PageRank and CheiRank probabilities and thus user balances are computed using the Google matrices $G$ and $G^*$ constructed from the global network of bitcoin transactions (\textit{a priori} no bankrupted users).
A user who went bankrupt remains in bankruptcy at all future iterations.
In this way we obtain the fraction, $W_c(\tau)=N_u(\tau)/N$, of users in bankruptcy or in crisis
at different iteration times $\tau$.

\section{Results}

\begin{figure}[!t]
	\centering
	\includegraphics[width=\textwidth]{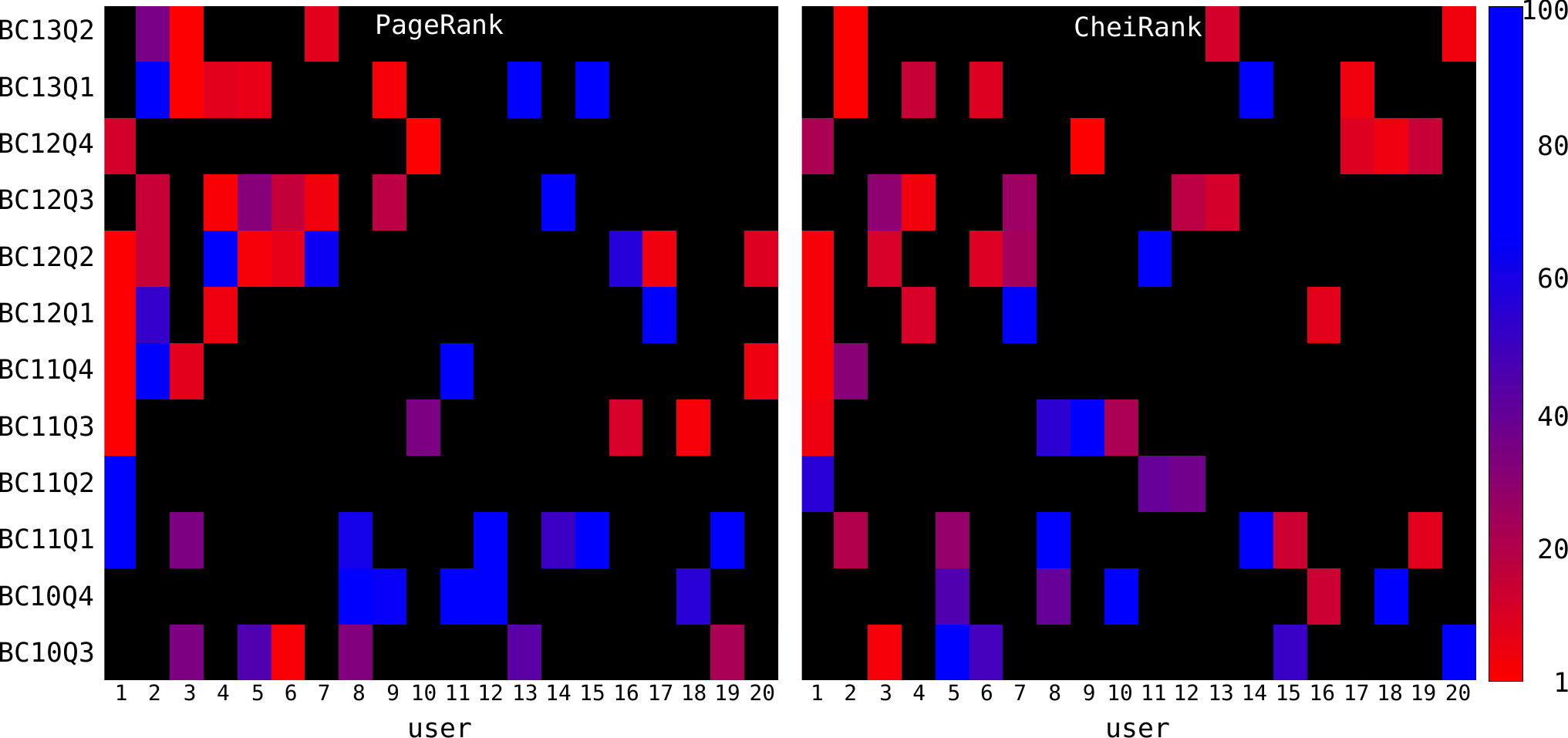}
	\caption{Twenty most present users in top100s of BCyyQq networks (see Tab.~\ref{tab1}) computed with PageRank (left panel) and CheiRank (right panel) algorithms. In horizontal axis the twenty users labeled from $1$ to $20$ are ranked according to the number of occurrences in the time slice top100s. The color ranges from red (user is ranked at the 1st position, $K=1$ or $K^*=1$) to blue (user is ranked at the 100th position, $K=100$ or $K^*=100$). Black color indicates a user absent from the top100 of the corresponding time slice.} \label{fig0}
\end{figure}

The PageRank and CheiRank algorithms have been applied to the bitcoin networks BCyyQq presented in Tab.~\ref{tab1}. An illustration showing the rank of the twenty most present users in the top 100s of these bitcoin networks is given in Fig.~\ref{fig0}. We observe that the most present user (\#1 in Fig.~\ref{fig0}) was, from the third quarter of 2011 to the fourth quarter of 2012, at the very top positions of both the PageRank ranking and of the CheiRank ranking. Consequently, this user was very central in the corresponding bitcoin networks with a very influential activity of bitcoin seller and buyer. Excepting the case of the most present user (\#1 in Fig.~\ref{fig0}), the other users are (depending of the year quarter considered) either top sellers (well ranked according to CheiRank algorithm, $K^*\sim1-100$) or top buyers of bitcoins (well ranked according to PageRank algorithm, $K\sim1-100$). In other words excepting the first column associated to user \#1 there is almost no overlap between left and right panels of Fig.~\ref{fig0}.

\begin{figure}[!th]
	\centering
	\includegraphics[width=\textwidth]{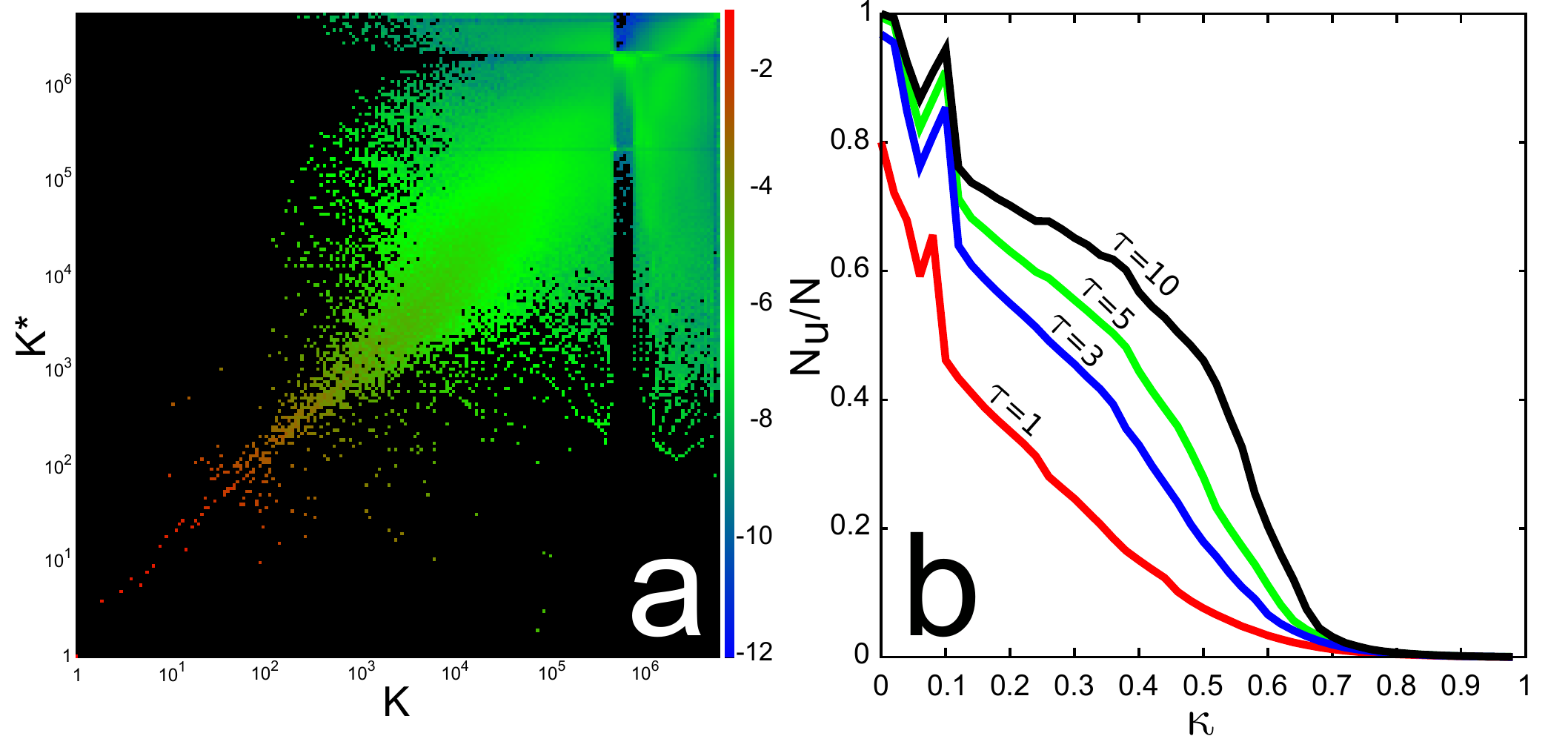}
	\caption{Panel a: density of users, $dN(K,K^*)/dKdK^*$, 
		in PageRank--CheiRank plane $(K,K^*)$ for BC13Q1  network;
		density is computed with $200 \times 200$ cells equidistant 
		in logarithmic scale; the colors are associated to the decimal logarithm 
		of the density; the color palette is a linear gradient from green color (low user densities) to red color (high user densities). Black color indicates absence of users.
		Panel b: fraction $N_u/N$ of BC13Q1 users in bankruptcy
		shown as a function of $\kappa$ for $\tau=1,3,5,$ and $10$.} \label{fig1}
\end{figure}

From now on we concentrate our study on the BC13Q1 network. For this bitcoin network,
the density of users on the PageRank-CheiRank plane $(K,K^*)$
is shown in Fig.~\ref{fig1}a. At low $K,K^*$, users are centered near
the diagonal $K=K^*$ that corresponds to the fact that on average users
try to keep balance between ingoing and outgoing bitcoin flows.
Similar effect has been seen also for world trade networks \cite{wtn1}. 

The dependence of the fraction of bankrupt users $W_c=N_u/N$
on the bankruptcy threshold $\kappa$ is shown in Fig.~\ref{fig1}b
at different iterations $\tau$. At low $\kappa < \kappa_c \approx 0.1$
almost 100\% of users went bankrupt at large $\tau=10$.

\begin{figure}[!t]
	\centering
	\includegraphics[width=\textwidth]{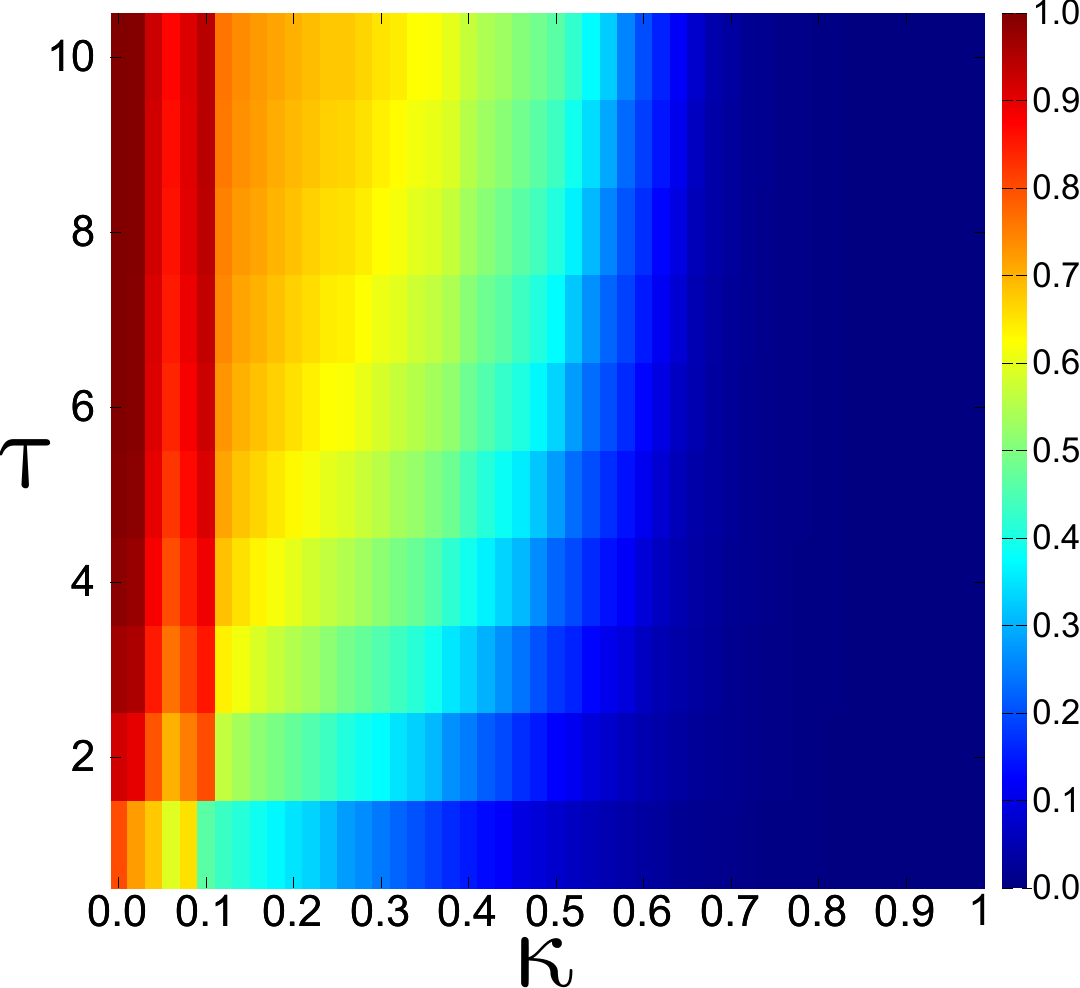}
	\caption{Fraction $N_u/N$ of BC13Q1 users in bankruptcy 
as a function of $\kappa$ and $\tau$.} \label{fig2}
\end{figure}

Indeed, Fig.~\ref{fig2} shows that the transition to bankruptcy
is similar to a phase transition so that at large $\tau$
we have $W_c=N_u/N \approx 1$ for $\kappa < \kappa_c \approx 0.1$,
in the range $\kappa_c\approx0.1 < \kappa < 0.55$ there are only
about $50\%$--$70\%$ of users in bankrupcy
while for $\kappa > 0.55$ almost all users
remain safe at large times.

\begin{figure}[!t]
	\centering
	\includegraphics[width=\textwidth]{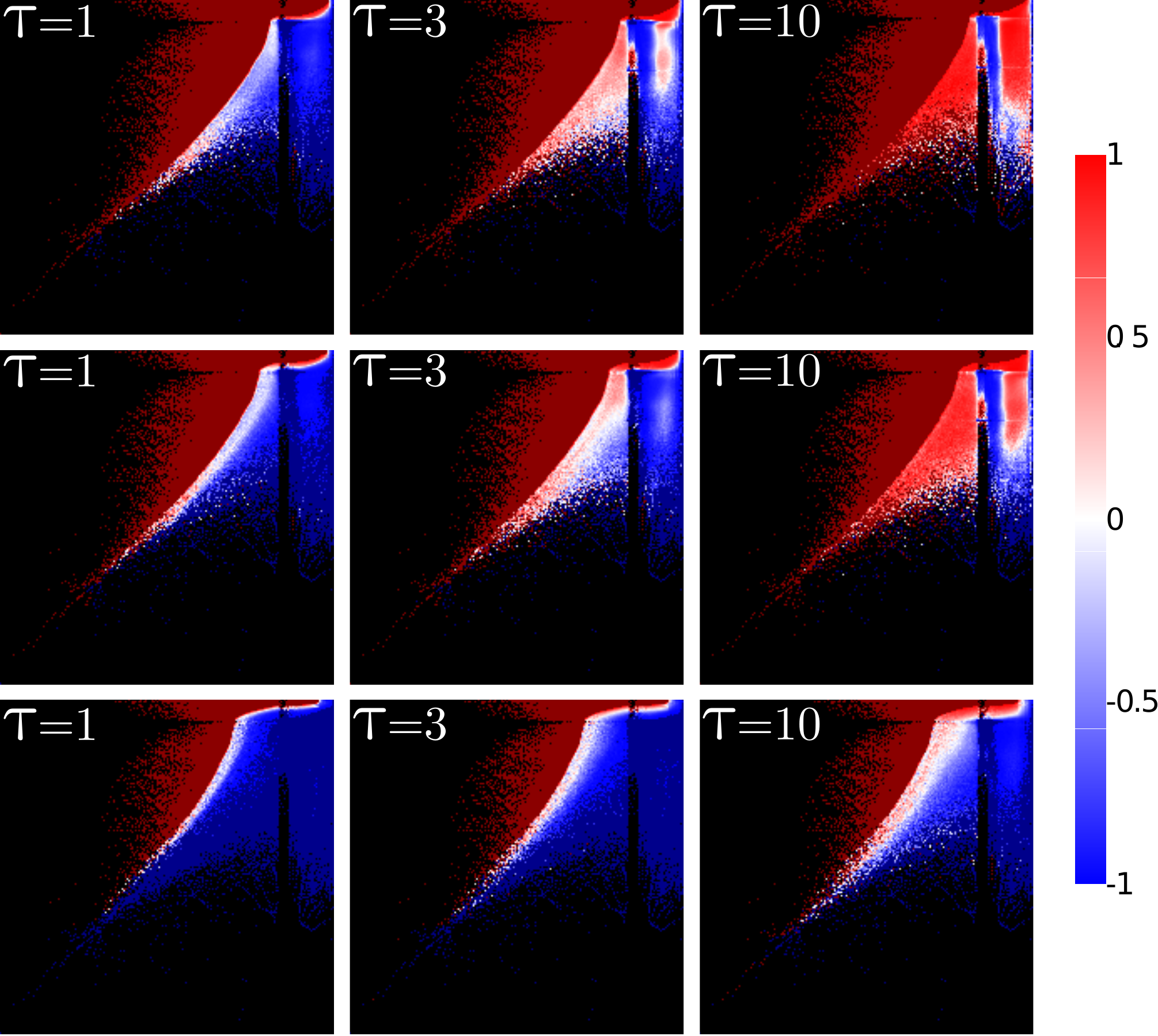}
	\caption{BC13Q1 users in bankruptcy (red) and safe (blue) 
		for $\kappa=0.15$ (top row), for $\kappa=0.3$ (middle row), and for $\kappa=0.6$ (bottom row). For each panel the horizontal (vertical) axis corresponds to PageRank (CheiRank) indexes $K$ $(K^*)$.
		In logarithmic scale, the $(K,K^*)$ plane has been divided in $200\times200$ cells. Defining $N_{\rm cell}$ as the total number of users in a given cell and $N_{u,\rm cell}$ as the number of users who went bankrupt in the cell until iteration $\tau$, we compute, for each cell, the value $(2N_{u,\rm cell}-N_{\rm cell})/N_{\rm cell}$ giving $+1$ if every user in the cell went bankrupt (dark red), $0$ if the number of users went bankrupt is equal to the number of safe users, and $-1$ if no user went bankrupt (dark blue). Black colored cells indicate cell without any user.} \label{fig3}
\end{figure}

The distribution of bankrupt and safe users on PageRank--CheiRank plane
$(K,K^*)$ is shown in Fig.~\ref{fig3} at different iteration times $\tau$.
For crisis thresholds $\kappa=0.15$ and $\kappa=0.3$, we see that very quickly users at top $K,K^* \sim 1$ indexes
go bankrupt and with growth of $\tau$ 
more and more users go bankrupt even if they  are located below
the diagonal $K=K^*$ thus having initially positive balance $B_u$.
However, the links with other users lead to propagation of
contagion so that even below the diagonal many users turn to
 bankruptcy.
This features are similar for $\kappa =0.15$
and $\kappa=0.3$ but of course the number of safe users is larger 
for  $\kappa=0.3$. For a crisis threshold $\kappa=0.6$, the picture is stable at every iterations $\tau$, the contagion is very moderate and concerns only the white region comprising roughly the same number of safe and bankrupt users. This white region broadens moderately as $\tau$ increases. We note that even some of the users above $K=K^*$ remain safe. We observe also that for $\kappa=0.6$ about a third of top $K,K^*\sim1$ users remain safe.

%
%
%
%
%
%
%
%
%
%
%
%

\begin{figure}[!t]
	\includegraphics[width=\textwidth]{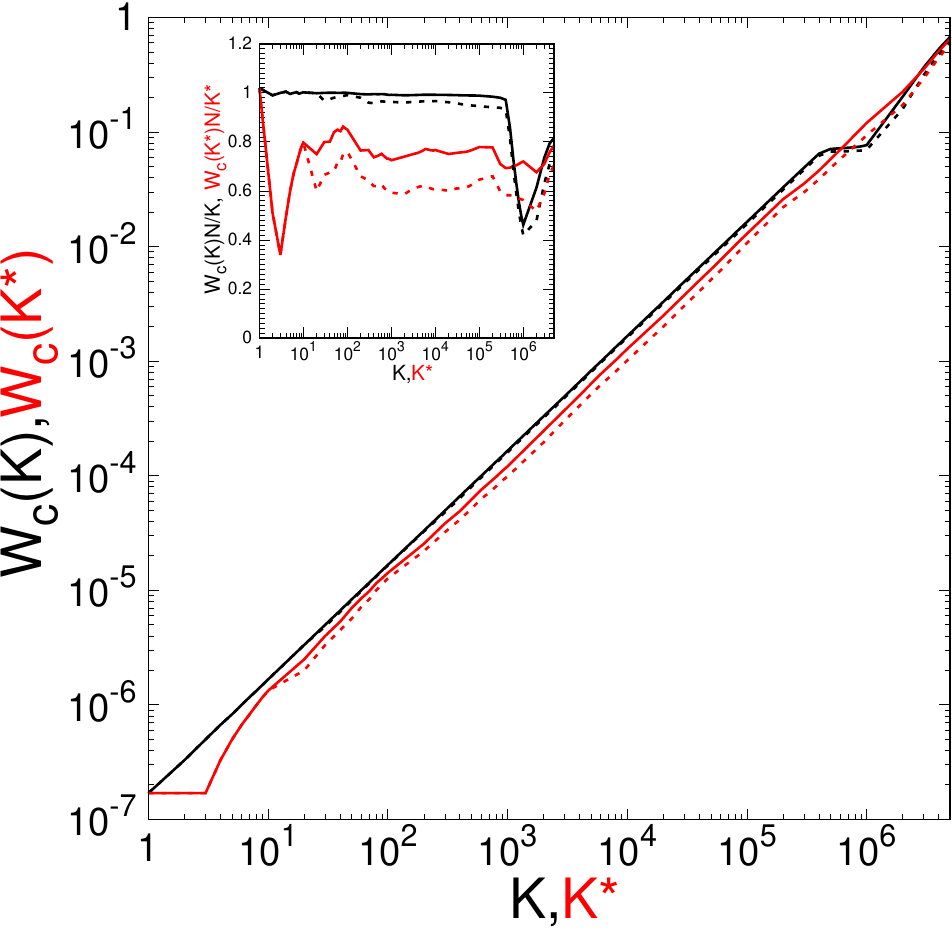}
	\caption{
		Integrated fractions, $W_c(K)$ and $W_c(K^*)$, of BC13Q1 users which went bankrupt at 
$\tau \leq 10$ for $\kappa=0.15$ (solid lines) and for $\kappa=0.3$ (dashed lines) as a function of PageRank index $K$ (black lines) and CheiRank index $K^*$ (red lines). The inset shows $W_c(K)N/K$ as a function of $K$ and $W_c(K^*)N/K^*$ as a function of $K^*$.
	} \label{fig4}
\end{figure}

Fig.~\ref{fig4} presents the integrated fraction, $W_c(K)=N_u(K)/N$, of users which have 
a Page\-Rank index below or equal to $K$ and which went bankrupt at $\tau\leq10$. 
We define in a similar manner the integrated fraction of CheiRank 
users $W_c(K^*)=N_u(K^*)/N$ being bankrupts. 
From Fig.~\ref{fig4} we observe $W(K)\approx K/N$ and $W(K^*)\approx K^*/N$.
Formal fits $W_c(K)=\mu^{-1}K^\beta$ of the data in the range $10 < K <10^5$ give $(\mu=5.94557\times10^6\pm95,\beta=0.998227\pm1\times10^{-6})$ for $\kappa=0.15$
and
$(\mu=5.65515\times10^6\pm231,\beta=0.99002\pm4\times10^{-6})$ for $\kappa=0.3$.
Formal fits $W_c(K^*)=\mu^{-1}K^{*\beta}$ of the data in the range $10 < K^* <10^5$ give $(\mu=1.03165\times10^7\pm3956,\beta=1.02511\pm3\times10^{-5})$ for $\kappa=0.15$
and
$(\mu=1.67775\times10^7\pm1.139\times10^4,\beta=1.05084\pm6\times10^{-5})$ for $\kappa=0.3$.

\begin{figure}[!t]
	\includegraphics[width=\textwidth]{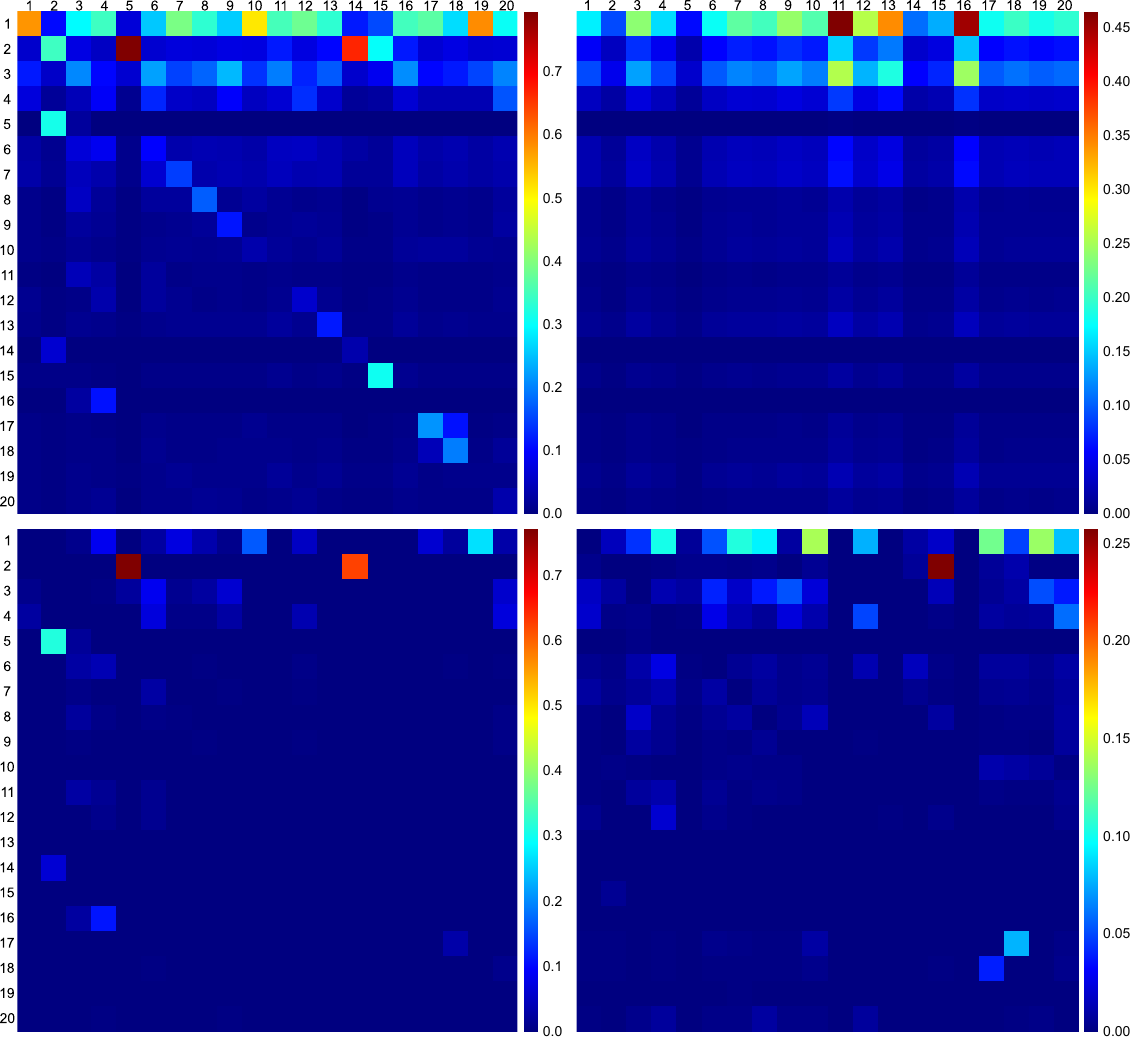}
	\caption{Reduced Google matrix $\GR$ associated to the top 20 PageRank users 
		of BC13Q1 network. The reduced Google matrix $\GR$ (top left) 
		has a weight $\WR=1$, its components $\Grr$ (bottom left), 
		$\Gpr$ (top right), and $\Gqrnd$ (bottom right) have weights 
		$\Wrr=0.29339$, $\Wpr=0.48193$, and $\Wqr=0.22468$ ($\Wqrnd=0.11095$). 
		Matrix entries are ordered according to BC13Q1 top 20 PageRank index.} \label{fig5}
\end{figure}

The results of contagion modeling show that PageRank and CheiRank top users $K,K^* \sim 1$
enter in contagion phase very rapidly.
We suppose that this happens due to strong interlinks existing between these users.
Thus it is interesting to see what are the effective links and interactions
between these top PageRank and top CheiRank users.
With this aim we construct the reduced Google matrix $\GR$
for the top 20 PageRank users of BC13Q1 network.
This matrix $\GR$ and its three components $\Gpr$, $\Grr$ and $\Gqrnd$
are shown in Fig.~\ref{fig5}.
We characterize each matrix component
by its weight defined as the sum of all matrix elements divided by $N_r =20$.
By definition the weight of $\GR$ is $\WR=1$.
The weights of all components are given in the caption of Fig.~\ref{fig5}.
We see that $\Wpr$ has the weight of about 50\% while $\Wrr$ and $\Wqr$ have 
the weight of about 25\%. These values are significantly higher comparing
to the cases of Wikipedia networks (see e.g. \cite{politwiki}).
The $\Grr$ matrix component (Fig.~\ref{fig5} bottom left panel) is similar to the bitcoin mass transfer matrix \cite{efsbitcoin} and the $(i,j)$-element of $\Grr$ is related to direct bitcoin transfer from user $j$ to user $i$. As $\Wrr=0.29339$, the PageRank top20 users directly transfer among them on average about $30\%$ of the total of bitcoins exchanged by these $20$ users. In particular, about $70\%$ of the bitcoin transfers from users $K=5$ and $K=14$ are directed toward user $K=2$. Also user $K=5$ buy about $30\%$ of the bitcoins sold by user $K=2$. We observe a closed loop between users $K=2$ and $K=5$ which highlights between them an active bitcoin trade during the period 2013 Q1. Also $30\%$ of bitcoins transferred from user $K=19$ were bought buy user $K=1$. The $20\times20$ reduced Google matrix $\GR$ (Fig.~\ref{fig5} top left panel) gives a synthetic picture of bitcoin direct and indirect transactions taking into account direct transactions between the $N\sim10^6$ users encoded in the global $N\times N$ Google matrix $G$. We clearly see that many bitcoin transfers converge toward user $K=1$ since this user is the most central in the bitcoin network. Although the $\Grr$ matrix component indicates that user $K=1$ obtains about $10\%$ to $30\%$ of the bitcoins transferred from its direct partners, the $\Gpr$ matrix component indicates that indirectly the effective amount transferred from direct and indirect partners are greater about $10\%$ to more than $45\%$. In particular, although no direct transfer exists from users $K=11$ and $K=16$ to user $K=1$, about $45\%$ of the bitcoins transferred in the network from users $K=11$ and $K=16$ converge indirectly to user $K=1$. Looking at the diagonal of the $\GR$ matrix we observe that about $60\%$ of the transferred bitcoins from user $K=1$ returns effectively to user $K=1$, the same happen, e.g, with user $K=2$ and user $K=15$ with about $30\%$ of transferred bitcoins going back. The $\Gqr$ matrix component (Fig.~\ref{fig5} bottom right panel) gives the interesting picture of hidden bitcoin transactions, i.e., transactions which are not encoded in the $\Grr$ matrix component since they are not direct transactions, and which are not captured by the $\Gpr$ matrix component as they do not necessarily involve transaction paths with the most central users. Here we clearly observe that $25\%$ of the total transferred bitcoins from user $K=15$ converge indirectly toward user $K=2$. We note that this indirect transfer is the result of many indirect transaction pathways involving many users other than the PageRank top20 users. We observe also a closed loop of hidden transactions between users $K=17$ and $K=18$.

\section{Discussion}

We performed the Google matrix analysis of Bitcoin networks
for transactions from the very start of bitcoins till
April 10, 2013. The transactions are divided by year quarters
and the Google matrix is constructed for each quarter.
We present the results for the first quarter of 2013
being typical for other quarters of 2011, 2012. 
We determine the PageRank and CheiRank vectors
of the Google matrices of direct and inverted bitcoin flows.
These probabilities characterize import (PageRank) and export (CheiRank)
exchange flows for each user (node) of the network.
In this way we obtain the dimensionless balance of each user $B_u$ 
($-1< B_u < 1$) and
model the contagion propagation on the network
assuming that a user goes bankrupt 
if its dimensional balance exceeds a certain bankruptcy threshold
$\kappa$ ($B_u \leq - \kappa$). We find that the phase transition takes place 
in a vicinity of the critical threshold $\kappa = \kappa_c \approx 0.1$
below which almost 100\% of users become bankrupts.
For $\kappa > 0.55$ almost all users remain safe
and for $0.1 < \kappa < 0.55$ about 60\% of users
go bankrupt. It is interesting that, as house of cards,
the almost all top PageRank and Cheirank users 
rapidly drop to bankruptcy
 even for $\kappa = 0.3$ being not very close to
the critical threshold $\kappa_c \approx 0.1$.
We attribute this effect to strong interconnectivity between
top users that makes them very vulnerable. 
Using the reduced Google matrix 
algorithm we determine the effective direct and indirect interactions
between the top 20 PageRank users
that shows their preferable interlinks including the long pathways via 
the global network of almost 6 million size.

We argue that the obtained results model the real situation of contagion
propagation of the financial and interbank networks.

{\it Acknowledgments:}
We thank L.Ermann for useful discussions.
This work was supported
by the French ``Investissements d'Ave\-nir'' program, 
project ISITE-BFC (contract ANR-15-IDEX-0003)
and
by the Bourgogne Franche-Comt\'e Region 2017-2020 APEX project (conventions
2017Y-06426, 2017Y-06413, 2017Y-07534; 
see \url{http://perso.utinam.cnrs.fr/~lages/apex/}).
The research of DLS is supported in part by  the Programme Investissements
d'Avenir ANR-11-IDEX-0002-02, 
reference ANR-10-LABX-0037-NEXT France (project THETRACOM).

%
%
%
%

\end{document}